\documentclass[twocolumn,twoside,slac_two]{revtex4}
\usepackage{graphicx}
\usepackage{fancyhdr}
\def\etal{{\it et al.}}

\begin{document}

\title{Imaging at Both Ends of the Spectrum: the Long Wavelength Array and Fermi}


\author{G.B. Taylor on behalf of the LWA Collaboration}
\affiliation{University of New Mexico}
%
%\author{P. Lucas}
%\affiliation{FNAL, Batavia, IL 60510, USA}

\begin{abstract}
The Long Wavelength Array (LWA) will be a new multi-purpose radio
telescope operating in the frequency range 10-88 MHz. Scientific
programs include pulsars, supernova remnants, general transient
searches, radio recombination lines, solar and Jupiter bursts,
investigations into the "dark ages" using redshifted hydrogen, and
ionospheric phenomena. Upon completion, LWA will consist of 53 phased
array “stations” distributed accross a region over 400 km in
diameter. Each station consists of 256 pairs of dipole-type antennas
whose signals are formed into beams, with outputs transported to a
central location for high-resolution aperture synthesis imaging. The
resulting image sensitivity is estimated to be a few mJy (5$\sigma$, 8
MHz, 2 polarizations, 1 h, zenith) from 20-80 MHz; with angular
resolution of a few arcseconds. Additional information is online at
http://lwa.unm.edu. Partners in the LWA project include LANL, JPL,
NRL, UNM, NMT, and Virginia Tech.

The full LWA will be a powerful instrument for the study of particle
acceleration mechanisms in AGN.  Even with the recently completed
first station of the LWA, called “LWA1”, we can begin spectral studies
of AGN radio lobes.  These can be combined with Fermi observations.
Furthermore we have an ongoing project to observe Crab Giant Pulses in
concert with Fermi.  In addition to these pointed studies, the LWA1
images the sky down to declination $-$30 degrees daily.  This is quite
complimentary to Fermi's daily images of the sky.
\end{abstract}

\maketitle

\thispagestyle{fancy}


\section{Introduction}

LWA1 originated as the first ``station'' (beamforming array) of the
Long Wavelength Array (LWA).  The LWA concept was conceived by 
Perley \& Erickson~\cite{perley} and expanded by 
Kassim \& Erickson~\cite{kas98} and Kassim \etal~\cite{kas05}.  It gained momentum with sub-arcminute
imaging with the VLA at 74 MHz~\cite{kas93}\cite{Kassim+07} and 
the project began in April 2007,
sponsored primarily by the Office of Naval Research (ONR), with the
ultimate goal of building an aperture synthesis radio telescope
consisting of 53 identical stations distributed over the
U.S. Southwest~\cite{elling09}.

\begin{figure*}[t!]
\begin{center}
\includegraphics[width=5.9in,angle=0]{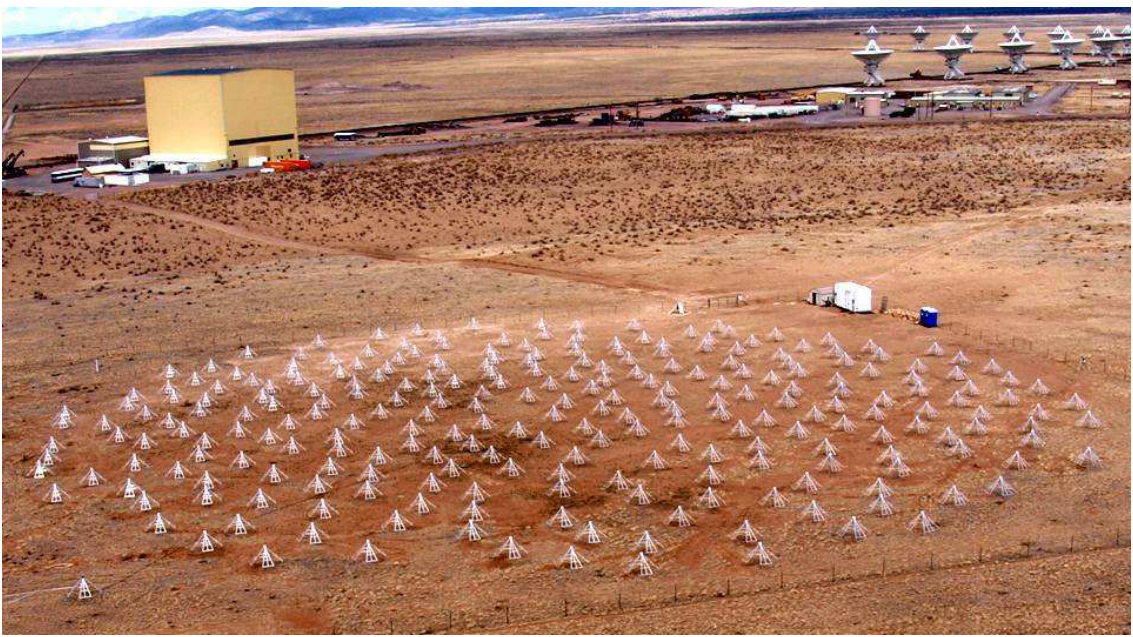}
\end{center}
\vspace{-0.5cm}
  \caption{
Aerial view of the LWA1 radio observatory. 
}
\label{fig:lwa_design}
\end{figure*}

The LWA1 Radio Observatory is shown in Fig.~\ref{fig:lwa_design}.  It
is located on NRAO land within the central square mile of the EVLA,
which offers numerous advantages.  The project to design and build
LWA1 was led by UNM, who also developed analog receivers and the
shelter and site infrastructure systems.  The system architecture was
developed by VT, who also developed LWA1's monitor \& control and data
recording systems.  Key elements of LWA1's design were guided by
experience gained from a prototype stub system project known as LWDA,
developed by NRL and the University of Texas at Austin~\cite{york07}; 
and by VT's Eight-meter wavelength Transient Array (ETA;~\cite{Deshpande09}).  NRL developed LWA1's active antennas, and JPL
developed LWA1's digital processing subsystem.  

Institutions represented in the LWA (as determined by attendance at
the May 12, 2011 LWA1 User Meeting) include U.S. Air Force Research
Laboratory (AFRL), Arizona State University (ASU), Harvard University,
Kansas University (KU), Long Island University, National Radio
Astronomy Observatory (NRAO), NASA Jet Propulsion Laboratory (JPL),
U.S. Naval Research Laboratory (NRL), New Mexico Tech (NMT),
University of New Mexico (UNM), UTB, and Virginia Tech (VT).  New
institutions and individuals are invited to join the LWA and if
interested should contact Namir Kassim (NRL) or Greg Taylor (UNM).
The LWA1 has recently been established as a University Radio
Observatory by NSF and as such will entertain regular calls for
proposals from the astronomical community.  The first of these widely
open calls for the proposals is out with a {\bf deadline of June 22, 2012}.
Table~1 summarizes the capabilities of LWA1.  For more details see the
LWA web pages at {\tt http://lwa.unm.edu} including the LWA Memo
series.

\begin{table*}[h!]
\begin{center}
\small
\caption{Summary of LWA1 Specifications}
\vspace{0.1cm}
\begin{tabular}{ll}
Specification & As Built Description \\
\hline 
\hline
\vspace{-0.2cm}\\
Beams: & 4, independently-steerable, each dual-polarization \\ 
Tunings: & 2 independent center frequencies per beam \\ 
Freq Range: & 24--87~MHz ($>$4:1 sky-noise dominated); 10--88~MHz usable \\ 
Instantaneous bandwidth: & $\le$16 MHz $\times$ 4 beams $\times$ 2 tunings \\ 
Minimum channel width: & $\sim$0 (No channelization before recording) \\ 
Beam FWHM: & [8,2]$^{\circ}$ at [20,80]~MHz for zenith-pointing \\ 
Beam SEFD: & $\sim$3~kJy (approximately frequency-independent) zenith-pointing \\ 
Beam Sensitivity: & $\sim5$~Jy (5$\sigma$, 1~s, 16~MHz) for zenith-pointing \\ 
All-Dipoles Modes: & TBN: 67~kHz bandwidth continuously from every dipole \\ 
& TBW: Full band (78~MHz) every dipole for 61~ms, once every $\sim$5 min.\\
\hline
\\
\end{tabular}
\end{center}
\end{table*}

\section{Current Status}

\begin{figure*}[t!]
\begin{center}
\includegraphics[width=5.0in,angle=0]{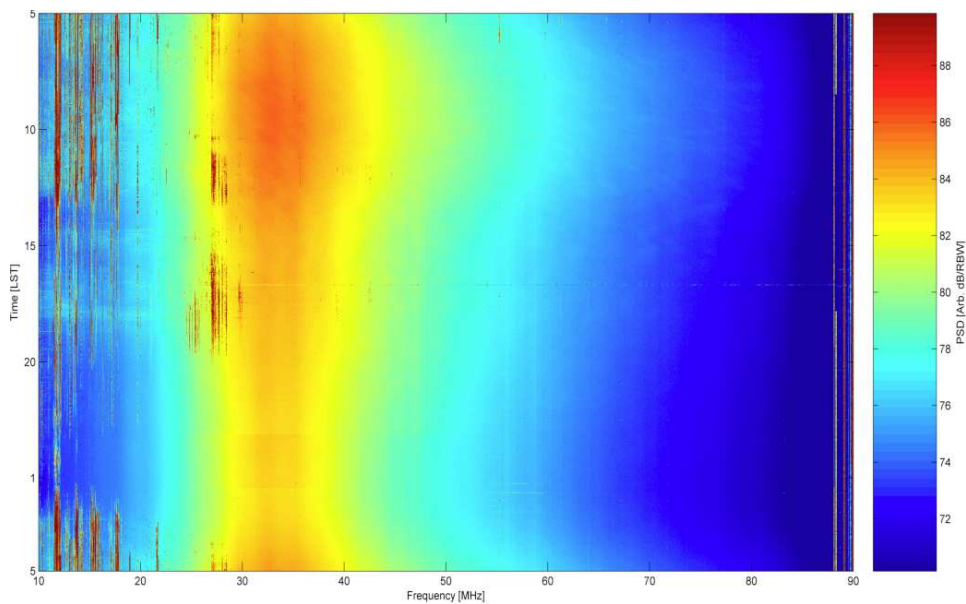}
\end{center}
\vspace{-0.5cm}
  \caption{
Spectrum using the TBW capture mode for 20 dipoles
phased at zenith for 24 hours.  The time and frequency variation of the 
background are real; the contribution of the active antenna appears
as a steep role-off below 30 MHz. Note that 30-88 MHz is always
useable, and even frequencies as low as 13 MHz are usable for 
a few hours each day.
}
\label{fig:TBW}
\end{figure*}

At the time of writing, we are in the commissioning phase.  We
anticipate to reach IOC (``initial operational capability'') --
essentially the beginning of routine operation as an observatory -- by
April 2012.  We now summarize some early results obtained during
commissioning.  In Fig.~\ref{fig:TBW} we show a spectrogram obtained
from the Transient Buffer Wideband (TBW) data taken over 24 hours for
a 20-dipole zenith-pointing beam.  The integration time of the
individual captures is 61 ms, and one capture was obtained every
minute.  The frequency resolution is $\sim$10 kHz and diurnal variation
of galactic noise is clearly evident.
Strong RFI from the FM bands shows up as vertical lines at 88
MHz and above.  Below 30 MHz there are a variety of strong
communications signals.  While there is abundant RFI visible in the
spectrum, it is very narrowband, obscures only a tiny fraction of our
band, and does not interfere with our ability to be sky-noise limited.
More details about the RFI envirornment can be found in Obenberger \&
Dowell~\cite{obenberger11}.

\begin{figure*}[t!]
\begin{center}
\includegraphics[width=5.2in,angle=0]{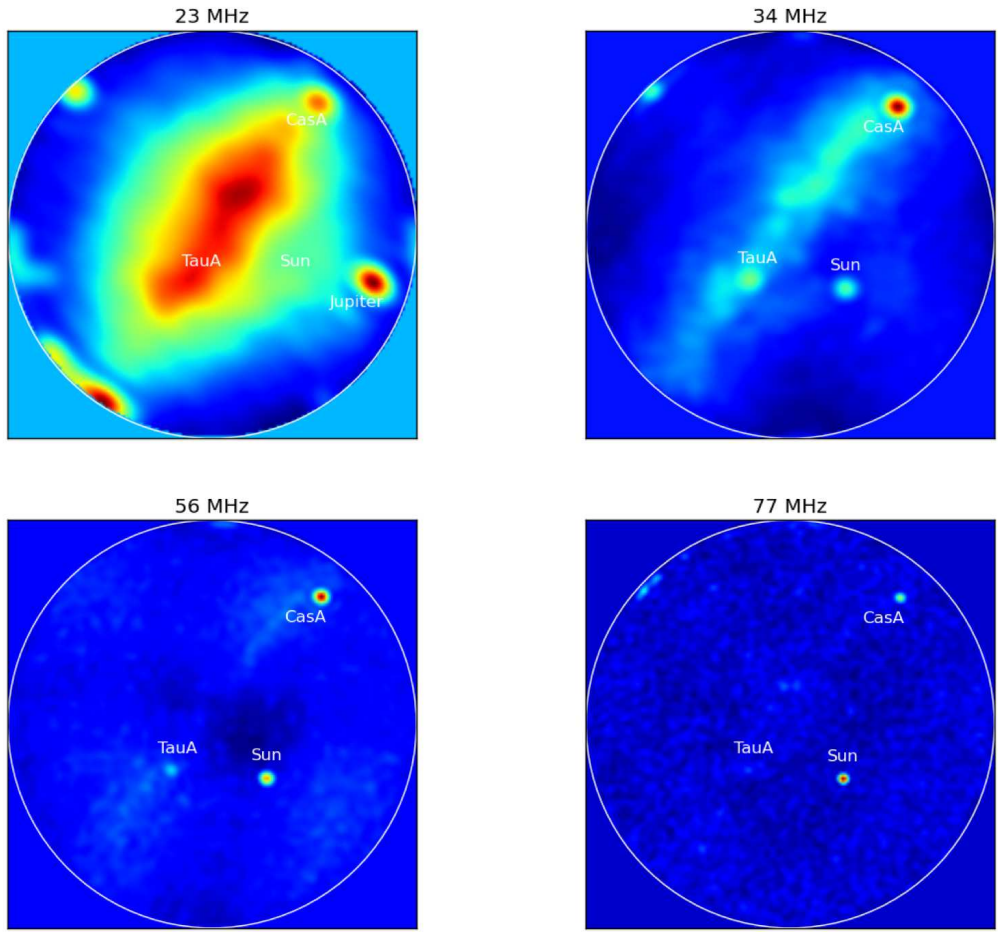}
\end{center}
\vspace{-0.5cm}
  \caption{
Nearly-simultaneous all-sky images taken at 4 widely separated frequencies
using LWA1's TBN mode.  Absolute calibration is the same in all four images;
the apparent decrease in sky brightness with increasing frequency is real, and
the bright region near zenith is the Galactic plane.  Clearly visible at 23 MHz
is Jupiter, and the horizon ``hot spots'' in the 23 MHz image  are ionospherically-refracted RFI.
Note that Cas~A and the Sun are visible in all images.  
Data was obtained for 10 seconds each, 50 kHz bandwidth, using 210 dipoles.  
}
\label{fig:all_sky}
\end{figure*}

We have recently begun imaging the sky with LWA1.
In Fig.~\ref{fig:all_sky} we show four views of the sky taken with 
the TBN mode on May 16 using 210 stands (21945 baselines).  In these 
Stokes I images one can see the
Galactic plane, Cas A, and Tau A, and at the lowest frequency 
Jupiter is quite prominent.  The LWA1 routinely images the sky
in near real-time using the Transient Buffer Narrowband (TBN) 
cabability of the station and a modest cluster located at the LWA1
(see Hartman et al. 2012, in preparation, for more details).
These images are shown live on ``LWA-TV'' which is available from
the LWA web pages.  Movies for each day are also available made
from the individual 5 second cqptures.

\section{Connections to Fermi}

\subsection{Pulsars}

\begin{figure*}[h!]
\begin{center}
\includegraphics[width=4.7in,angle=-90]{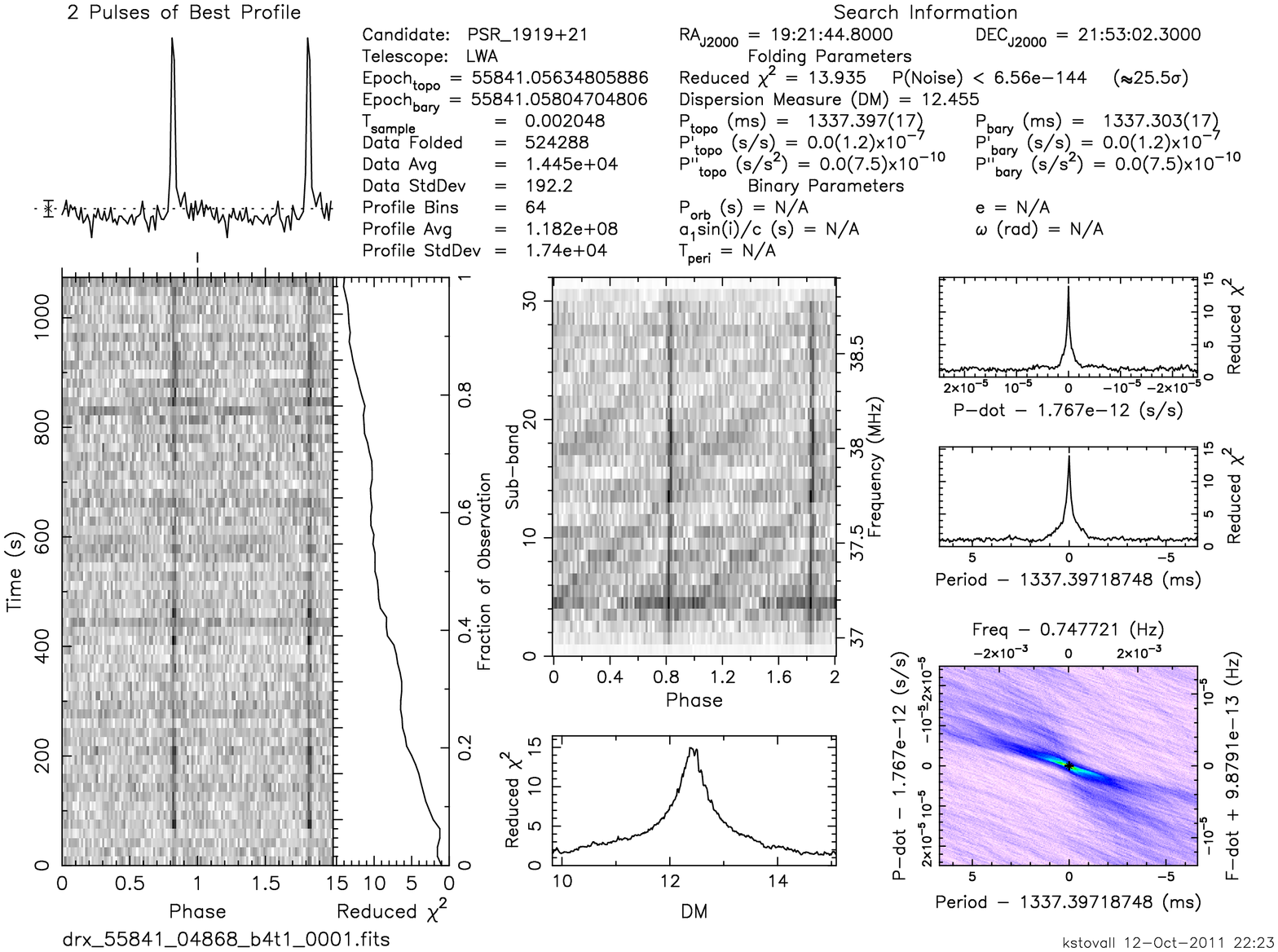}
\end{center}
\vspace{-0.5cm}
  \caption{
First detection with LWA1 of the pulsar 1919+21 using a narrow (1 MHz) beam 
taken during commissioning observations.
}
\label{fig:B1919+21}
\end{figure*}

Pulsars are fascinating objects with spin periods and magnetic fields
strengths ranging over 4 and 5 orders of magnitude respectively.
Though it is well accepted that pulsars are rotating neutron stars,
the pulsar emission mechanism and the geometry of the emitting region
are still poorly understood~\cite{Eilek05}.  LWA1 will be an
excellent telescope for the study of pulsars including single pulse
studies, and studies of the interstellar medium (ISM).  In fact, it is
in the LWA1 frequency band range where strong evolution in pulsar
radio emission can be observed, e.g, a turn over in the flux density
spectrum, significant pulse broadening, and complex changes in the
pulse profile morphologies~\cite{mal94}. And, although
pulsars were discovered at low frequency (81 MHz), 
there is a remarkable lack of good
observational data in the LWA1 frequency range. LWA1 observations
should be able to detect dozens of pulsars (e.g.,
Fig.~\ref{fig:B1919+21} and see~\cite{Jacoby}) in less than 1000
seconds.

LWA1 will be able to perform spectral studies of
pulsars over a wide frequency range and with high spectral resolution.
This will allow investigators to look for drifting subpulses.  
Strong notches have been seen to appear in the profiles of 
pulsars at low frequencies~\cite{tim}, but little progress
has been made in understanding their origin.
Some pulsars may reach 100\% linear polarization at low frequencies
(B1929+10;~\cite{man73}).
In addition to being intrinsically of interest (providing clues
about the pulsar magnetospheric structure), such strongly 
polarized beacons can assist in probing coronal mass ejections
and determining the orientation of their magnetic fields, which
strongly affects their impact on Earth.  

LWA1's large collecting area will be particularly useful for ``single pulse''
science, including studies of Crab Giant Pulses (CGPs) and Anomalously
Intense Pulses (AIPs).
The Crab Pulsar intermittently produces single pulses having intensity
greater than those of the normal periodic emission by orders of
magnitude. Despite extensive observations and study, the mechanism
behind CGPs remains mysterious. 
Observations of the Crab pulsar across the electromagnetic
spectrum can distinguish between various models for GP emission such as
enhanced pair cascades, radio coherence, and changes in beaming direction.
We plan to coordinate low frequency observations of GPs with observations
of GPs by Fermi.
To date the study of the CGP emission at low radio
frequencies is only very sparsely explored.
Reported modern observations of CGPs in this frequency
regime are limited to 
just a few in recent years including UTR-2 at 23 MHz~\cite{popov},
MWA at 200 MHz~\cite{bhat}, and LOFAR LBA~\cite{stap11}.
LWA1 will be able to provide hundreds of hours per year of sensitive
observations of CGPs which will revolutionize our knowledge of the 
time- and frequency-domain characteristics of these enigmatic events.
Combined with observations in other wavelength regimes (e.g., simultaneous
L-band observations already planned in a current observing project)
significant advances in understanding are expected.

We should be able to measure scattering for practically every good
CGP detection (S/N $\sim$20 or better), and it is known that both the
dispersion and scattering of the Crab emission can vary dramatically
on short or long time scales. By observing over an extremely
broad bandwidth, we may be able to better quantify the scatter
broadening and thereby assess the level and importance of
anisotropy. Furthermore, the broad bandwidth of the observations will
be helpful in shedding light on the issue of the frequency scaling of
the scattering (believed to be $\sim$3.6 compared to the canonical value
of $\sim$4.4 for the general ISM), which is thought to be related to the
nature of turbulence in the nebula.

Anomalous high intensity single pulses from known pulsars have been
reported previously using the UTR-2 (Ulyanov \etal\ 2006) and
LOFAR~\cite{stap11}.  These anomalously intense pulses (AIPs) have
many features similar to the giant pulse phenomenon, including
emission in a narrow longitude interval and power-law distribution of
the pulse energy.  One distinctive feature of these AIPs, however, is
that they are generated by subpulses or some more short lived
structures within subpulses.  The emission is seen to be quite narrow
band, typically 1 MHz in bandwidth.  The nature of such pulses is not
yet understood.  The LWA1 with its excellent sensitivity and large
available bandwidth provides an opportunity to study these pulses.

Pulsars make up a significant component of the source population
visible by both Fermi and the LWA1.  The LWA1 is an excellent 
instrument for the study of pulsars as it offers good
sensitivity, broad bandwidth, wide field-of-view for rapid 
survey speed, and precise timing capabilities.  The LWA1 also 
records raw voltages, which allows for very flexible post-processing
of the data (coherent dedispersion, fine time and frequency resolution, etc.).
See Fig.~\ref{fig:B1919+21} for the first light detection of pulsar 1919+21.

An immediate connection between LWA1 and Fermi is in the study of 
pulsars.  In particular Fermi has recently discovered over 36 pulsars
in the gamma-ray band.  Only a few of these have been found to 
pulse in the radio at centimeter wavelengths.  Low frequency 
searches are of considerable interest as pulsars are generally 
very steep spectrum and the beaming fraction is
thought to be lower at low frequencies.  A survey of gamma-ray
pulsars will be carried out with the LWA1 in the near future.

\subsection{Blazars}

At high galactic latitudes 80\% (106 of 132) of the $\gamma$-ray
bright sources detected in the LAT Bright Source List (BSL) derived
from the first 3 months of Fermi observations~\cite{abd09} are
associated with known Active Galactic Nuclei (AGN).  In the second
LAT AGN catalog (2LAC; \cite{ack11}) there are 1017 $\gamma$-ray sources
associated with AGN.  The vast majority
of the Fermi $\gamma$-ray sources are blazars, with strong, compact
radio emission.  These blazars exhibit flat radio spectra, rapid
variability, compact cores with one-sided parsec-scale jets, and
superluminal motion in the jets~\cite{mar06}.  Extensive studies of
blazars are reported in these proceedings.

Due to the low angular resolution of LWA1, only a few blazars are bright 
enough to rise above the confusion noise.  Fortunately, blazars are
in general highly variable so it is possible to detect flaring sources
with the LWA at low frequency.  The all-sky images from the 
Prototype All Sky Imager (PASI) on LWA1 (see \S3.3) can 
be compared to the daily all-sky images from Fermi.  Strong flaring
blazars can be detected in the all-sky images, and beams can be used
to confirm detections.  Measurements at low frequencies can help
to constrain the particle acceleration mechanisms is the jets.

\subsection{Transients}

Astrophysical transient sources of radio emission signal the explosive
release of energy from compact sources (see Lazio \etal\ 2010, Cordes
\& McLaughlin 2003 for reviews).  Known types of radio transients
include cosmic ray airshowers, solar flares (\S2.5), Jovian flares and
flares from extrasolar hot jupiters (\S2.2), giant flares from
magnetars (Cameron \etal\ 2005), rotating radio transients (McLaughlin
\etal\ 2006), giant pulses from the Crab pulsar, and supernovae.  The
study of these sudden releases of energy allow us to recognize
these rare objects, and yield insights to the nature of the
sources including energetics, rotation rates, magnetic field
strengths, and particle acceleration mechanisms.  Furthermore, some
radio transients remain unidentified such as the galactic center radio
transient GCRT J1745$-$3009 (Hyman \etal\ 2005), and require further study. 

PASI is a software correlator and
imager for LWA1 that analyzes 
continuous samples from all dipoles with a 75~kHz passband placeable
anywhere within 10--88~MHz.  PASI images nearly the
whole sky ($\approx$$1.5\pi$~sr) every five seconds, continuously
and in near realtime, with full Stokes parameters and typical sensitivities
of $\sim$5~Jy at frequencies above 40~MHz and $\sim$20~Jy at 20~MHz.
Candidate detections can be followed up within seconds by beamformed
observations for improved sensitivity and localization.  These
capabilities provide an unprecedented opportunity to search the
synoptic low-frequency sky.  PASI 
saves visibility data for $\sim$20 days, allowing it to ``look back
in time'' in response to transient alerts. The images generated
by PASI will be archived indefinitely.  The images of the sky from
PASI are available "live" at the URL:
{\tt http://www.phys.unm.edu/~lwa/lwatv.html}.

\bigskip % extra skip inserted

\begin{acknowledgments}

  Construction of the LWA has been supported by the Office of Naval
  Research under Contract N00014-07-C-0147. Support for operations and
  continuing development of the LWA1 is provided by the National
  Science Foundation under grant AST-1139974 of the University Radio
  Observatory program.

\end{acknowledgments}

\bigskip % extra skip inserted

\end{document}